# Similarity Analysis of Blood Count Reference Intervals Across Continents Reveals No Reproducible Population or Geography-Linked Structure and Supports Personalised Values

**Kunlin Wu**[1], **Abicumaran Uthamacumaran**[2], **Hector Zenil**[1,2,3*]

[1]Algorithmic Dynamics Lab, King's Institute for Artificial Intelligence, King's College London, London, UK

[2] Oxford Immune Algorithmics, Oxford University Innovation & London Institute for Healthcare Engineering, Reading, UK

[3] Departments of Biomedical Computing and Digital Twins, School of Biomedical Engineering and Medical Sciences, King's Faculty of Life Sciences and Medicine, King's Health Partners AHSC (NHS–KCL), King's College London, London, UK

**Abstract**

Blood reference intervals (RIs) underpin diagnostic interpretation and therapeutic monitoring worldwide. However, many widely used RI systems originate from limited historical cohorts and have been propagated across health systems without harmonised derivation protocols, shared metadata, or cross-population validation. Consequently, the current global RI landscape reflects a heterogeneous mixture of legacy standards and local laboratory practices rather than a biologically grounded global framework. Here we examine the structure of published Complete Blood Count (CBC) reference intervals, one of the most commonly used laboratory panels worldwide. We compiled CBC RI data from 28 countries and analysed their similarity using a multi-stage framework combining variability mapping, hierarchical clustering, information-theoretic distances, cohesion benchmarking, and nonlinear manifold visualisation. Body mass index (BMI) served as a methodological positive benchmark and exhibited clear continent-level clustering (mean cohesion ≈ 0.78–0.81). In contrast, CBC reference intervals showed no reproducible geography-linked clustering across analytical methods, with uniformly high cohesion scores (mean ≈ 1.27–1.30). Weak signals in red-cell indices (MCV, HGB) were unstable across sexes and distance metrics. This absence of structure should not be interpreted as evidence that current CBC reference intervals represent universal biological standards. Rather, it is more consistent with the fragmented and historically inherited nature of the global RI landscape. These findings indicate that currently published CBC reference intervals do not encode coherent global structure and therefore provide limited support for universal population-based diagnostic thresholds. Instead, they support a transition toward recalibrated and personalised reference frameworks based on longitudinal individual baselines and harmonised derivation standards.

* Corresponding author: hector.zenil@kcl.ac.uk

## Introduction

Reference intervals (RIs) are widely applied in clinical laboratory testing as a standard of care for disease screening, diagnosis, and therapeutic monitoring [1–2]. Their establishment directly influences diagnostic sensitivity and specificity and can determine whether a patient is classified as physiologically normal or abnormal. However, the global reference-interval landscape has not emerged from a unified or harmonised derivation framework. Instead, many currently used RIs originate from historically local datasets, laboratory-specific conventions, and legacy clinical standards that were subsequently propagated across health systems without consistent recalibration or cross-population validation. As a result, widely used RI systems often represent heterogeneous historical artefacts rather than biologically comparable global standards.

This study evaluates the structure of published reference-interval systems themselves rather than individual-level hematologic biology, asking whether currently reported RIs encode reproducible cross-population organisation (i.e., geographic structure) or instead reflect heterogeneous derivation practices across laboratories and health systems. RIs originate from statistical summaries and lack fine-grained individual characterisation [3]. Studies (e.g., NHANES) show CBC differences by race/ethnicity for HGB, WBC, and PLT, implying that "universal" RIs may not be globally applicable [4]. A harmonization view (e.g., the International Federation of Clinical Chemistry and Laboratory Medicine, IFCC) promotes unified international standards to improve comparability [5], whereas others warn uniform RIs can obscure risk signals and exacerbate health inequities in heterogeneous populations [6-7].

Machine learning and deep clustering have been explored to model latent variability [8-9], but without standardized QC or multinational sources these remain experimental [10]. Some argue population RIs are insufficient for precision medicine and advocate individual reference values (RCVs; reference change value) or longitudinal trajectories [11], consistent with integrating genomic, environmental, lifestyle, and social data for individualized prevention/treatment [12-13]. Yet RI systems still rely on Western, European-descent data, lacking cross-national/ethnic/geographical modeling, leading to reduced accuracy and inequity, especially in developing regions and minority populations [7,14-15]. Body Mass Index (BMI) shows systematic ethnic and regional variation [16], suggesting even "universal" indicators can be population specific. Herein, "geography" is defined strictly as continent- or region-level grouping of countries, used as a spatial proxy for shared environmental and health ecosystem context. "Ethnicity" or genetic ancestry is not directly measured or inferred, and any references to ethnicity are discussed in the context of prior literature rather than modelled variables.

These limitations raise a critical methodological question: Do currently published reference intervals encode any coherent and reproducible population-level structure at all? If reference intervals are derived from biologically grounded and comparable population samples, one would expect some degree of reproducible geographic or demographic organisation to emerge when they are analysed across countries. Conversely, if the global RI landscape is primarily shaped by heterogeneous laboratory practices, historical conventions, and inconsistent derivation methods, then published reference intervals should fail to exhibit stable cross-population structure.

By analysing CBC reference intervals from 28 countries, we aim to evaluate whether currently published RI systems encode reproducible population structure or instead reflect heterogeneous clinical derivation practices. Importantly, this study examines the structure of the reference-interval system itself rather than underlying individual hematologic biology. A failure to detect reproducible geographic organisation would therefore suggest that the global RI landscape is dominated by methodological and historical variability rather than physiologically meaningful population structure. Establishing this distinction is essential for guiding the transition from legacy population-based reference systems toward recalibrated and ultimately personalised diagnostic frameworks.

We focus on four gaps: (1) regional bias in data sources (NHANES [18], Japan [19], NORIP [20]) with limited representation from Africa, Middle East, South Asia, Latin America [5,7,21,22] raising equity concerns [23]; (2) coarse granularity due to summary-level data (intervals or mean±2SD) that mask intra-population heterogeneity and lack longitudinal tracking [24-27]; (3) methodological homogeneity with limited structural modeling and little analyte-level discriminability quantification (mutual information, permutation importance, random forest) limiting interpretability and decision support; (4) theoretical homogeneity assuming universality, despite arguments for context-specific, dynamically adjusted values and ongoing RCV debates without a bridge from population structure to individual baselines [5-7,28]. Response strategy. Build a cross-continental CBC RI database; combine variability mapping, clustering, and interpretable dimensionality reduction; compare linkage–distance combinations; quantify analyte contributions; and propose a pathway toward geo-ethnicity-adaptive and ultimately individualised RIs.

Accordingly, failure to detect geography-linked organisation in published RIs should not be interpreted as evidence that universal reference intervals are biologically appropriate; rather, it may reflect the heterogeneous and historically inherited nature of the current RI landscape.

## Methods

The aims of this study were to compile a multinational CBC reference interval (RI) database and incorporate BMI as a positive benchmark case; quantify cross-country variability in CBC using visualization and SDR; apply multiple clustering strategies to both CBC and BMI and benchmark their geographical sensitivity using the Two-Level Cohesion Score; use UMAP with feature-importance metrics to evaluate whether geography contributes to latent structure in CBC versus BMI; and, finally, develop a methodological framework that integrates variability mapping, clustering evaluation, and interpretable dimensionality reduction as a foundation for future geography-adaptive RI systems.

## Data Collection

**CBC:** We first sought official national RI websites; however, most countries lacked standardized CBC RIs. Reasons include the absence of unified national standards, intra-country regional variation [29], and policies granting local autonomy (e.g., Dubai Health Authority) [30]. We therefore adopted a diversified sourcing strategy: 12 countries from peer-reviewed RI studies; 9 from university-affiliated, teaching, or public hospitals; 3 from local health authorities; 1 from a textbook; 1 from medical-school hematology lecture notes; 1 from a private laboratory; and 1 from an unofficial document attributed to an authoritative hospital. Where authoritative sources were absent, we substituted with geographically and demographically similar neighboring countries; for geographically unique cases (e.g., Greece), we used the most credible available source. Source type and confidence level were annotated for transparency. The final dataset comprised 28 countries. We collected RIs for WBC, Platelets, MCHC, MCV, MCH, RBC, HGB, and HT. Sweden discontinued routine MCHC reporting as of 2015-10-14 (Karolinska), and MCHC values were therefore unavailable for this country [31].

For analytes such as WBC and platelets, many authoritative hospital standards report a single reference interval intended for use across sexes. These values were therefore retained as sex-agnostic, consistent with their clinical application. No imputation across sexes was performed. No substitution or averaging across sexes was applied. The number of such sex-undifferentiated reference intervals is reported in Supplementary Table S5.

Hospital-derived RIs for WBC, Platelets, MCHC, MCV, and MCH were typically not sex-specific, whereas academic sources were generally sex-stratified. When sex-specific RIs were missing for analytes that are commonly stratified (e.g., HGB, HT, RBC in Canada and the Philippines), equivalence was assumed to reflect published clinical standards. Country-level source details are provided in Appendix A.

Information on analytical platforms, reagents, and laboratory instrumentation was not available for most countries and could not be consistently retrieved from authoritative sources. CBC reference intervals across countries may derive from different analytical platforms (e.g., Sysmex, Abbott, Mindray) with distinct calibration procedures and analytical sensitivities. As such, no analyser-level adjustment was performed. This limitation reflects the absence of standardized reporting in current reference-interval publications rather than a methodological oversight. To mitigate avoidable variability, analyses were restricted to adult reference intervals and to the most recent authoritative version available for each country.

**BMI (positive case):** We obtained mean BMI values for the same 28 countries from the WHO Global Health Observatory [32] for both sexes across six age groups (60–64, 65–69, 70–74, 75–79, 80–84, 85+), yielding 12 sex–age strata to mirror MI-based top-5 feature selection. Although the most recent data year was 2016, BMI was used solely as a methodological benchmark rather than for temporal surveillance.

## Data Analysis

All analyses and figures used Python frameworks. Comparative visualization with standard deviation ratios (SDR). For each analyte/sex, we visualized RIs and computed relative SDR (coefficient of variation) by taking country-level midpoints, then the mean and standard deviation (SD) of midpoints, and SD/mean as SDR; each country was annotated with source type and confidence level 1–3. To assess whether conclusions depended on representing reference intervals by midpoints, we performed parallel analyses using lower limits, upper limits, and interval widths (upper–lower). All analyses employed identical clustering, cohesion, and dimensionality-reduction procedures. This allowed direct comparison of results across alternative representations of reference intervals

**Phylogenetic trees:** We constructed univariate (per analyte, by sex) and multivariate (all eight analytes, by sex) trees. We benchmarked diverse linkage–distance pairs to test robustness across dependence structures: Euclidean for magnitude-based geometry, Manhattan for L1 robustness, Cosine/Correlation for orientation and shape similarity, and Mutual Information (MI) for nonlinear and distributional dependencies, providing complementary sensitivity profiles across analytes and sexes [33] (See Table 1). We compared Ward and Average linkage via a hierarchical, distance-based clustering algorithm, Unweighted Pair Group Method with Arithmetic Mean (UPGMA) [33]. For univariate Average linkage, Euclidean and Manhattan yielded identical structures; MI/Cosine/Correlation are not applicable to univariate. Multivariate trees used Ward+Euclidean; Average with Euclidean, Cosine, Manhattan, Correlation; and Average+MI. MI distance required feature selection by MI to “Continent”: top-5 features (and a top-3 sensitivity) to balance noise vs coverage. Only three analytes had nonzero MI; two zero-score analytes were added to reach five, which still altered dendrograms due to their contribution to global pairwise distances. We repeated clustering with top-3 only to assess stability. MI discretization used quantile binning with five bins; alternative binning would change outcomes.

All dendrograms represent hierarchical clustering of countries based on z-score–standardized reference-interval representations (midpoints, bounds, or widths), using specified distance metrics and linkage rules as stated in figure captions. Ward linkage with Euclidean distance was used for CBC analyses as a variance-minimizing baseline suited to continuous, low-signal reference-interval data, providing the most stable and conservative clustering. In contrast, Average linkage with Mutual Information distance was used for BMI as a positive benchmark to maximize sensitivity to nonlinear, population-structured dependencies, where geography-linked signal is known to exist.

**Robinson–Foulds tree-distance analysis.**
Robinson-Foulds (RF) tree distances (scores) were measured between dendrograms generated from the alternative linkage–distance metrics to characterise their topological convergence. RF distances were computed using skbio.tree.TreeNode.compare_rfd (unrooted; proportion = True) across all CBC reference-interval representations (midpoints, lower bounds, widths) for both sexes with N_PERM = 1000. The same standardized input matrices and linkage–distance settings used in our main findings were repeated. RF distances were derived separately for male and female CBC RI representations and compared across midpoint, lower-bound, upper-bound, and width-based trees. Lower RF scores indicate greater topological concordance between clustering methods, while higher scores denote instability.

**Mutual Information Estimation and Binning Sensitivity.**
Mutual information (MI) was computed using a histogram-based estimator applied to discretized analyte values. Hereafter, ‘MI-based’ refers to analyses using mutual information–derived distance or feature selection. To evaluate robustness to discretization choices, MI-based two-level cohesion scores were systematically recomputed across a wide range of bin counts (3–20 bins). Analyses were performed using fixed feature selections (top-3 and top-5 MI-ranked analytes), and results were evaluated separately for lower bounds, midpoints, upper bounds, and interval widths. While absolute MI values varied with bin number, the qualitative absence of stable geography-linked clustering remained consistent across all binning configurations, sexes, and reference-interval representations (Supplementary Figures S1–S2).

Since MI estimation depends on discretization, absolute MI magnitudes and cohesion values are not directly comparable across different binning schemes. Accordingly, our conclusions rely on invariance of qualitative clustering outcomes rather than on any single numerical MI value.

**Two-Level Cohesion Score:** For each dendrogram, we computed mean cophenetic distance across within-continent country pairs, then averaged across continents; lower scores indicate stronger geographical cohesion. We summarized scores across all linkage–distance combinations for univariate and multivariate analyses. We computed matrix-level statistics including Mantel tests (10 000 permutations), Pearson/Spearman concordance between distance and similarity matrices, Receiver Operating Characteristic – Area Under the Curve (ROC-AUC) for discriminative association, and MDS stress/variance explained to quantify geometric coherence. We also computed Cohen's *d* effect size to quantify standardized mean separation between high- and low-correlation pairs, and an Odds Ratio (Fisher's exact test) comparing the likelihood of strong correlations among short- versus long-distance pairs to provide interpretable effect-size validation. Additionally, we computed distance-correlation (dCor) between the flattened upper triangles of the distance matrix and *(1–correlation)* to capture nonlinear concordance, and quantified variance partitioning by the eigenvalue spectrum of the classical Multidimensional Scaling (MDS; Gower-centred) matrix (reporting cumulative variance for 2D–5D). MDS is a dimensionality-reduction technique that represents pairwise distances between samples in a low-dimensional feature space, preserving relative distance relationships in a global structure.

**UMAP & feature importance:** We embedded male and female CBC data with Uniform Manifold Approximation and Projection (UMAP) to visualize high-dimensional structure, then quantified analyte contributions via absolute Pearson correlations with UMAP1/UMAP2 (averaged), permutation importance (projection sensitivity), and random forest importance (Gini impurity reduction). We applied the same pipeline to BMI for a positive-case benchmark. All UMAP embeddings were generated using the umap-learn Python implementation with fixed parameter settings (n_neighbors = 5, min_dist = 0.3, metric = 'Euclidean' distance) and a fixed random seed (random_state = 42) to ensure reproducibility of the reported figures. Despite fixed seeds, UMAP remains a stochastic algorithm and small variations in embedding geometry can occur across runs and environments. All downstream analyses were therefore interpreted at the level of qualitative structure and stability rather than exact coordinate positions. Permutation-based importance scores depend on the learned UMAP embedding and therefore inherit its stochasticity. As a result, absolute importance values may vary slightly across runs, while relative analyte rankings and qualitative conclusions remain stable. As a stochastic method, UMAP preserves local neighborhood structure. 95% bootstrapped Confidence Intervals (CI) (N_BOOT = 500) were added to the fitted regression/trends. While represented as confidence ellipses in the Main text, box plot representations are provided in the Supplementary Information.

| Linkage Method | Distance Metric | Applicability | Sensitivity | Rationale |
|---|---|---|---|---|
| **Ward** | Euclidean | Univariate + Multivariate | Absolute magnitude differences | Baseline metric for variance-minimizing hierarchical structure |
| **Average (UPGMA)** | Euclidean | Univariate + Multivariate | Absolute magnitude differences | Benchmark against Ward to assess linkage dependence |
| **Average (UPGMA)** | Manhattan | Multivariate only | Robustness to large deviations | Captures additive (L1) structure and reduces influence of outliers |
| **Average (UPGMA)** | Cosine | Multivariate only | Directional similarity | Tests whether countries cluster by relative analyte *profiles* rather than raw levels |
| **Average (UPGMA)** | Correlation | Multivariate only | Shape/orientation of co-fluctuations | Detects similarity in analyte *patterns* independent of absolute values |
| **Average (UPGMA)** | Mutual Information | Multivariate only | Nonlinear/dependence structures | Captures non-Euclidean similarity potentially missed by linear metrics |

**Table 1.** Phylogenetic Tree clustering algorithms. Overview of linkage–distance combinations evaluated for phylogenetic modelling of CBC reference intervals. Euclidean is applicable to both univariate and multivariate analyses, while Cosine, Correlation, Manhattan, and Mutual Information distances are applied only to multivariate feature spaces.

**Monte Carlo Sensitivity Analysis**

To assess the sensitivity of the analytical framework to continent-level effects, we performed a simulation-based power analysis using permutation testing. Analyses were conducted on CBC RI midpoints from 28 countries, stratified by sex and analyte. Only analyte–sex combinations with sufficient representation (≥2 continents, ≥2 countries per continent, ≥8 countries total) were included.

For each analyte, synthetic continent-level shifts ($\Delta$ = 0–2 pooled standard deviations, SD) were introduced by adding a fixed offset to one continent while preserving the empirical variance structure. Gaussian noise proportional to the pooled SD (10%) was added to reflect measurement heterogeneity. For each $\Delta$, 100 Monte Carlo simulations were performed. Detection was evaluated using a permutation-based omnibus test across continents. A one-way ANOVA F-statistic was used, with significance assessed via 100 random permutations of continent labels to obtain empirical p-values ($\alpha = 0.05$). Power was defined as the proportion of simulations yielding significant permutation p-values. All simulations were implemented in Python (NumPy, Pandas, SciPy) with fixed random seeds to ensure reproducibility.

## Results

### Cross-national variability in CBC reference intervals

Figure 1A–D shows the between-country distribution of WBC and MCV RIs, stratified by sex and annotated with data-source confidence levels. Both analytes display modest overall dispersion and no reproducible continent-level structure, and male-female patterns are concordant across countries. The SDR analysis confirms that between-population variability is low relative to within-population variance, indicating that these parameters are not geographically partitioned at the reference-interval level. Consistent results were obtained when tree and UMAP analyses were repeated using lower limits, upper limits, and interval widths, indicating that the lack of reproducible geography-level structure does not depend on midpoint-only representations

Other CBC analytes (RBC, HGB, HCT, MCH, MCHC, Platelets; additional panels in Figure 1) similarly exhibit heterogeneous but non-geographic scatter, where wider ranges in select countries appear attributable to local laboratory conventions, not regionally conserved physiology. Missing MCHC values for Sweden reflect a documented change in institutional reporting policy rather than biological divergence. Collectively, Figure 1 suggests that CBC reference intervals do not exhibit intrinsic geography-linked stratification, motivating the subsequent multivariate clustering analyses. Importantly, these observations pertain to the structure of available/published reference intervals themselves and do not directly characterise biological variability at the individual or population level.

To directly examine the structure of published reference intervals beyond midpoint representations, we first visualized complete lower–midpoint–upper reference ranges for a representative analyte (male white blood cell count) (Figure 2). Across 28 countries, reference interval bounds exhibited substantial overlap across continents, with low between-country dispersion (relative SDR = 0.105). No continent-level stratification was apparent at the level of raw reference intervals, motivating subsequent multivariate analyses.

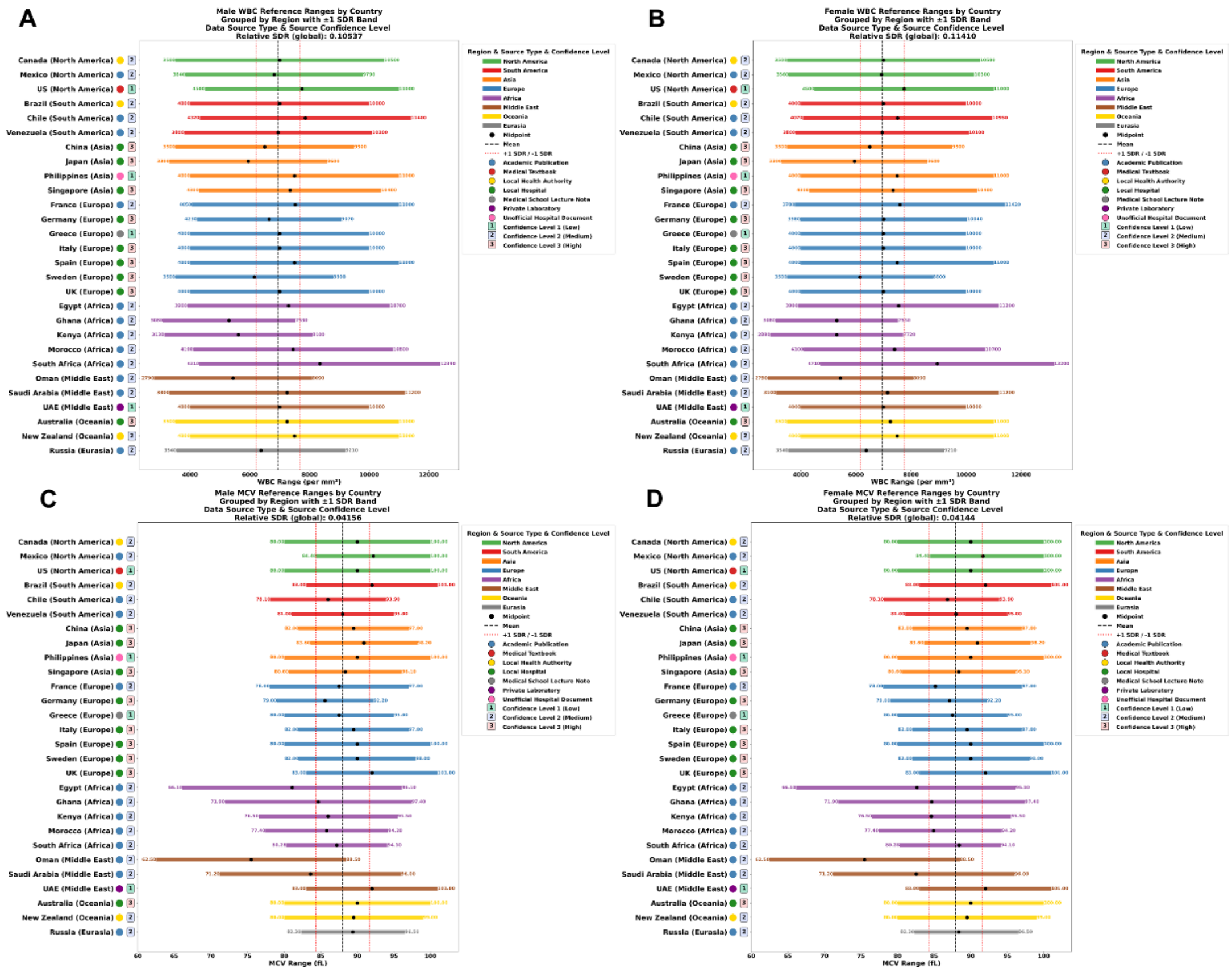

**Figure 1.** A) Male WBC Reference Ranges by Country and Region, with Data Source Type and Confidence Level**.** B) Female WBC Reference Ranges by Country and Region, with Data Source Type and Confidence Level**.** C) Male MCV Reference Ranges by Country and Region, with Data Source Type and Confidence Level**.** D) Female MCV Reference Ranges by Country and Region, with Data Source Type and Confidence Level.

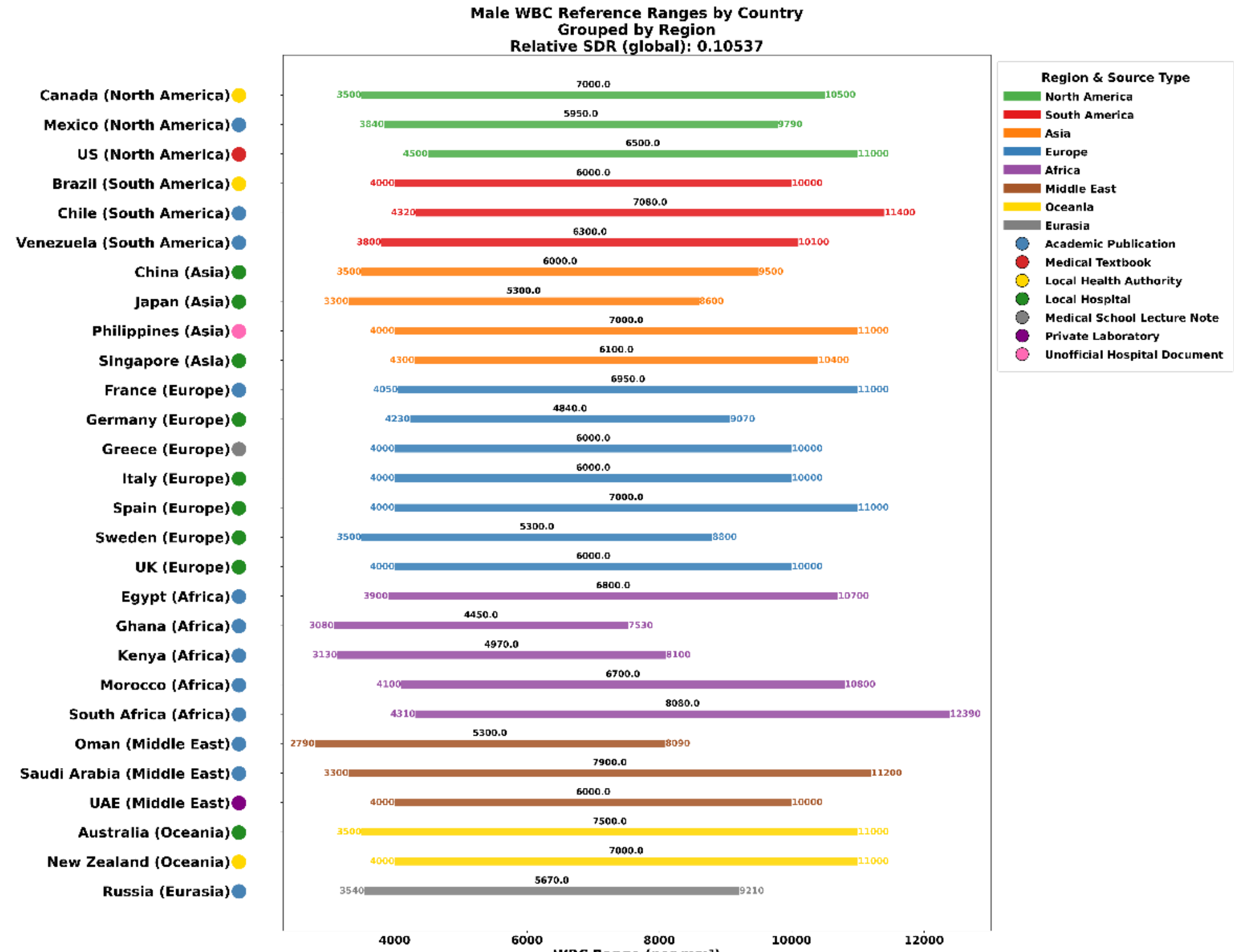

**Figure 2. Cross-national comparison of male white blood cell (WBC) reference intervals as a representative analyte.** Available lower limits, midpoints, and upper limits of male WBC reference intervals are shown for 28 countries, grouped by geographic region and annotated by source type. Despite methodological heterogeneity across sources, reference interval bounds show extensive overlap across continents, with low between-country dispersion (relative SDR = 0.105). This indicates that geographic structure is not evident at the level of raw reference intervals for this representative analyte.

### Multivariate clustering of CBC versus BMI

Figure 3A–B presents multianalyte phylogenetic clustering of CBC reference intervals by sex. Across all tested linkage–distance combinations, no persistent grouping of countries by continent is observed. Occasional local neighbourhood effects collapse when the metric or feature set is altered, indicating instability of any apparent CBC geography signal. Additional dendrograms for individual CBC analytes and alternative linkage–distance specifications are provided in the Supplementary Information. Mutual-information clustering accentuates subtle structure but fails to reproduce after re-specification (top-5 vs top-3 features), indicating that such structure is method-dependent rather than biologically anchored.

In contrast, BMI clustering in Figure 3C–D forms continent-coherent groupings under the same modelling framework, particularly with MI and correlation distances. Unlike CBC reference intervals, BMI values exhibited clear continent-level clustering, supporting the validity of the analytical framework. However, it is important to note that differences between BMI and CBC may partially reflect greater measurement heterogeneity in CBC reference intervals, which are derived from diverse clinical practices and publication contexts, rather than intrinsic physiological homogeneity. This establishes BMI as a positive control demonstrating that the pipeline detects geography-linked structure when present. The contrast between CBC and BMI directly supports the interpretation that CBC lacks stable geo-physiological boundaries, whereas BMI retains them.

Mutual-information feature ranking in the BMI trees (Figure 3C–D) showed that the geography signal was driven by mid- to late-adulthood BMI intervals: for males, the highest MI scores were observed for the 60–64 and 65–69 age groups (0.551 and 0.512), while for females the strongest contributions came from the 70–74 and 65–69 groups (0.581 and 0.531). This age-anchored concentration of MI signal indicates that continent-level clustering in BMI is not noise-driven but biologically structured, emerging in life-stages when adiposity reflects long-term regional lifestyle and environmental exposures rather than transient physiological fluctuation. In contrast, no analogous concentration or replicable “driver analyte” pattern was observed for CBC, explaining the failure of MI to stabilise continent-level groupings in CBC despite its strong performance on BMI.

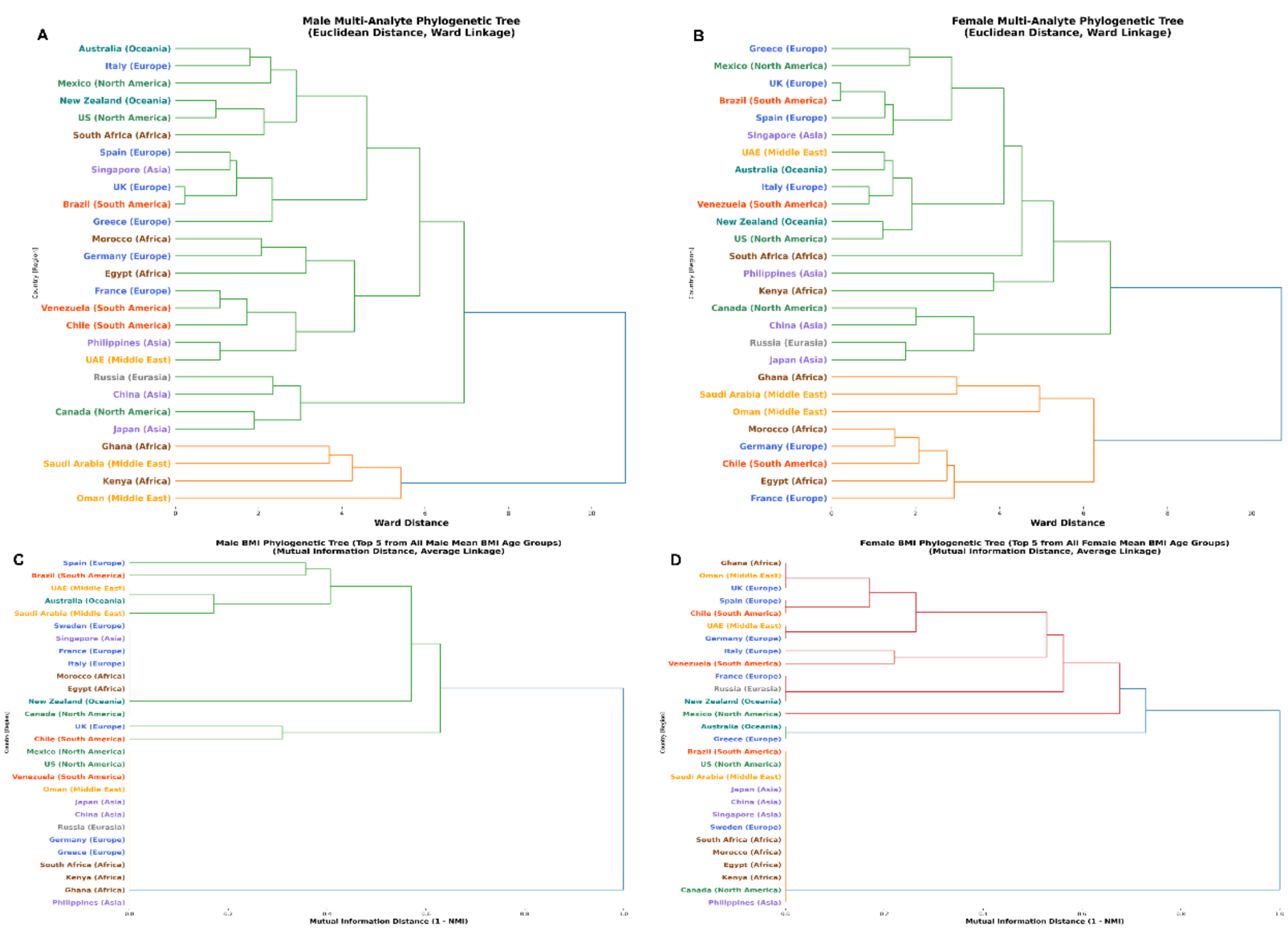

**Figure 3.** A) Male Multi-Analyte Phylogenetic Tree (Ward Linkage + Euclidean Distance). B) Female Multi-Analyte Phylogenetic Tree (Ward Linkage + Euclidean Distance). C) Multidimensional Phylogenetic Tree of Countries by All Male Mean BMI Age Groups (Average Linkage + Mutual Information Distance (Top 5)). D) Multidimensional Phylogenetic Tree of Countries by All Female Mean BMI Age Groups (Average Linkage + Mutual Information Distance (Top 5)). All dendrograms represent hierarchical clustering of countries based on z-

score–standardized reference-interval representations (midpoints, bounds, or widths), using specified distance metrics and linkage rules. Ward+Euclidean was selected as a conservative, variance-minimizing baseline for CBC, while Average+Mutual Information was used for BMI to maximize sensitivity to nonlinear, geography-linked population structure.

**Cohesion benchmarking**

To quantify the strength of continent-level organisation in each clustering approach, Two-Level Cohesion Scores were computed (Tables S1–S5 in Supplementary Information). For male and female CBC (Tables S1-S3), cohesion scores remained uniformly high across linkage–distance combinations, confirming weak within-continent similarity and the absence of reproducible geography-linked structure. Average + MI marginally lowered cohesion but did not stabilise any continent-level pattern, consistent with only weak and non-replicating geography effects in CBC. Bin-sensitivity analyses confirmed that while absolute cohesion magnitudes depend on discretization, the absence of stable geography-linked clustering is preserved across bin counts, sexes, and reference-interval representations (Supplementary Figures S1–S2). Observed midpoint-based cohesion scores and permutation-based analyte importance values are reported in Supplementary Tables S6 and S7, respectively.

In contrast, male and female BMI (Tables S4 and S5) exhibited substantially lower cohesion scores, particularly under information-theoretic and correlation-based distances. This concordance between topology (Figure 3) and cohesion benchmarking (Tables S2–S5) demonstrates that BMI encodes genuine geographic structure, whereas the null result for CBC is a bounded negative finding rather than a methodological limitation. Complementary effect-size analyses further supported this matrix-level consistency. High-correlation pairs were substantially closer in distance space (Cohen's $d = 1.24$), and pairs with short distances were ≈ 6-fold more likely to exhibit strong correlation (Odds Ratio = 6.15, $p < 0.0001$) confirming that the observed CBC null pattern reflects limited detectable structure in published RI data rather than measurement noise.

Quantitatively, mean cohesion scores across all non-Ward clustering strategies were 1.274 (male CBC) and 1.300 (female CBC), compared with 0.777 (male BMI) and 0.811 (female BMI), corresponding to an absolute ~0.49–0.50 reduction when geography genuinely contributes to structure. This corresponds to an approximate 38–42% relative reduction in within-continent cophenetic distance for BMI compared with CBC, representing a medium-to-large structural effect size. Notably, mutual-information distance ranked first in three of the four tables, but only yielded sustained low cohesion in the BMI case, confirming that MI functions as a true detector of structure rather than a false-positive amplifier. These findings quantitatively corroborate the dendrogram analyses and suggest that the absence of CBC clustering reflects a structural non-association in published reference-interval data.

To further assess concordance between the distance and similarity structures, we computed additional global metrics on two 103 × 103 matrices summarizing all pairwise relationships among national analyte reference datasets: one capturing inter-country dissimilarity (distance) and the other profile similarity (correlation). Because dendrograms were used for exploratory visualization rather than hierarchical inference, reproducibility was assessed using robust matrix-and geometry-based concordance metrics rather than node-level bootstrapping The Mantel test yielded *$r = 0.48$, $p = 0.0001$*, indicating strong correspondence between the two matrices, i.e., the pairwise distance geometry and correlation structure. Pearson and Spearman coefficients (both $\approx 0.48$, $p < 0.0001$) confirmed this linear and rank-level agreement. The discriminative accuracy of distance for identifying highly correlated pairs (area under the curve; AUC = 0.86) and the 2D classical multidimensional scaling (MDS) variance explained (64 %) further demonstrate that the matrix topology is internally consistent and geometrically stable.

A nonlinear robustness check using the Székely–Rizzo distance-correlation (dCor) test yielded a high concordance between distance and similarity matrices (dCor = 0.74, $p < 0.001$). Variance partitioning based on the eigenvalue spectrum of the classical MDS analysis showed that 3D, 4D, and 5D components explained 78%, 87%, and 92% of total relational variance, respectively, indicating a low-dimensional, geometrically coherent structure. This further validates that the observed absence of continent-level CBC clustering is consistent with the lack of a detectable population-level structure in accessible reference-interval systems rather than an artifact of modelling or metric choice.

## UMAP embeddings and feature importance

Figure 4A–B shows UMAP embeddings of CBC values for males and females. Country positions overlap extensively, and no consistent continent-separated patterns emerge. Feature-importance scoring (correlation, permutation, random-forest) identifies red-cell indices (most often MCV, sometimes HGB) as relatively more influential for local embedding geometry, but these signals remain non-geographic and sex-inconsistent, again confirming the absence of continent-level stratification. We emphasize that UMAP is used here as a qualitative manifold visualization and stress-test for latent structure, rather than as a deterministic geometric model; conclusions are therefore based on reproducible overlap patterns and robustness across representations, not on exact embedding coordinates. Conclusions were not conditioned on exact UMAP coordinates, and qualitative embedding patterns were interpreted only in conjunction with distance-based and cohesion analyses.

BMI clustering dendrograms in Figure 3C–D show clearer continent-coherent groupings, and the low cohesion scores in Tables S4–S5 quantitatively corroborate this topology. This further validates that the analytic workflow detects geography-linked structure when it is present in the data.

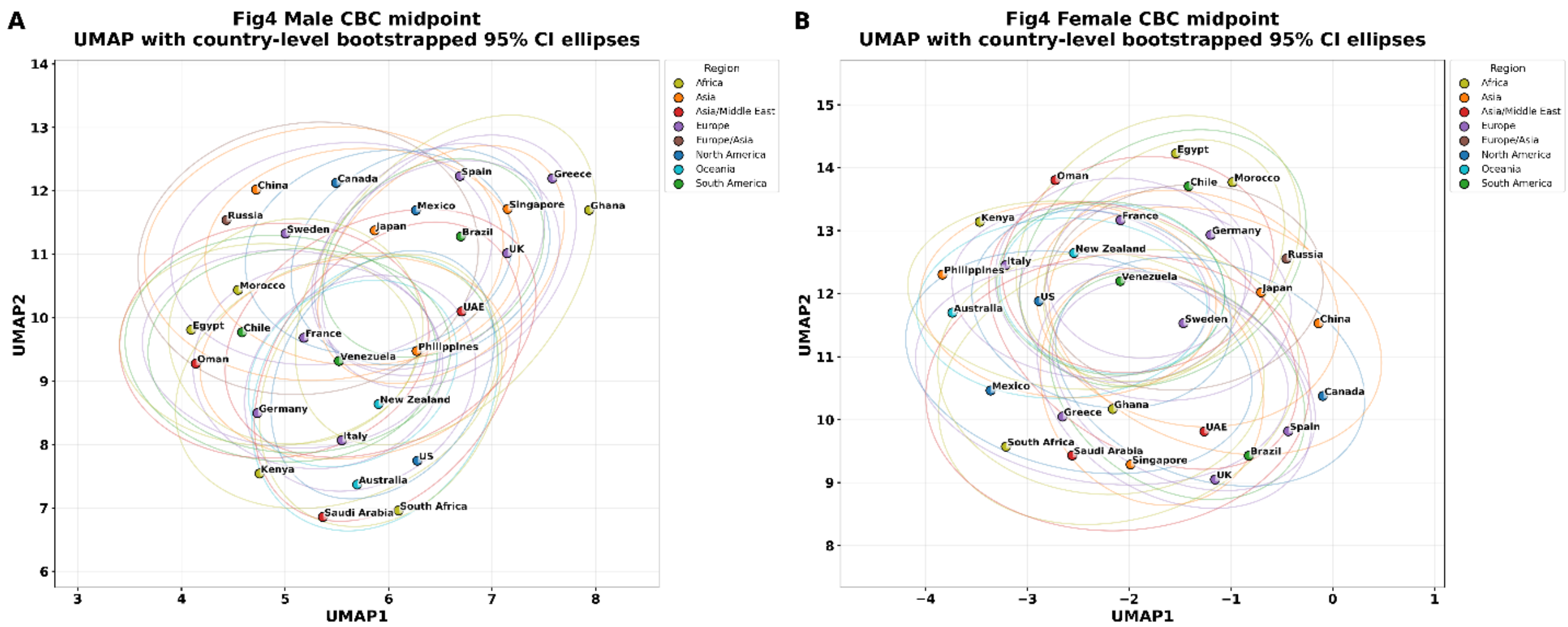

**Figure 4. UMAP visualization of midpoint-based CBC reference intervals with regional 95% confidence ellipses.** A) UMAP Visualization and Correlation Score of All Male CBC Analytes. B) UMAP Visualization and Correlation Score of All Female CBC Analytes. The bootstrapped 95% confidence ellipses summarize regional dispersion in UMAP space and show substantial overlap across regions, indicating no reproducible continent-level separation.

To address the concern that midpoint-only representations may obscure population-level variability, we repeated the full multivariate analysis using lower reference limits and interval widths (upper–lower) for both males and females. As shown in Figure 5, UMAP embeddings derived from these alternative representations similarly failed to reveal stable continent-level structure. This consistency across reference-interval representations and sex indicates that the observed lack of geographic clustering does not depend on midpoint-based simplification.

Regional 95% confidence ellipses derived from bootstrapped UMAP embeddings showed substantial overlap across all continental groupings in both midpoint-based (Figure 4) and alternative representation limits (Figure 5), with no reproducible continent-level separation. Feature-importance analyses across these representations further revealed that no single analyte consistently dominated permutation importance scores across both sexes and representations, and rank orderings shifted substantially between lower-bound and interval-width panels (Table S8). This representation-dependence of importance scores confirms that apparent analyte contributions to UMAP geometry reflect methodological sensitivity rather than stable biological signal. This reinforces that the absence of geography-linked CBC structure is robust to nonlinear embedding regardless of how RIs are summarised.

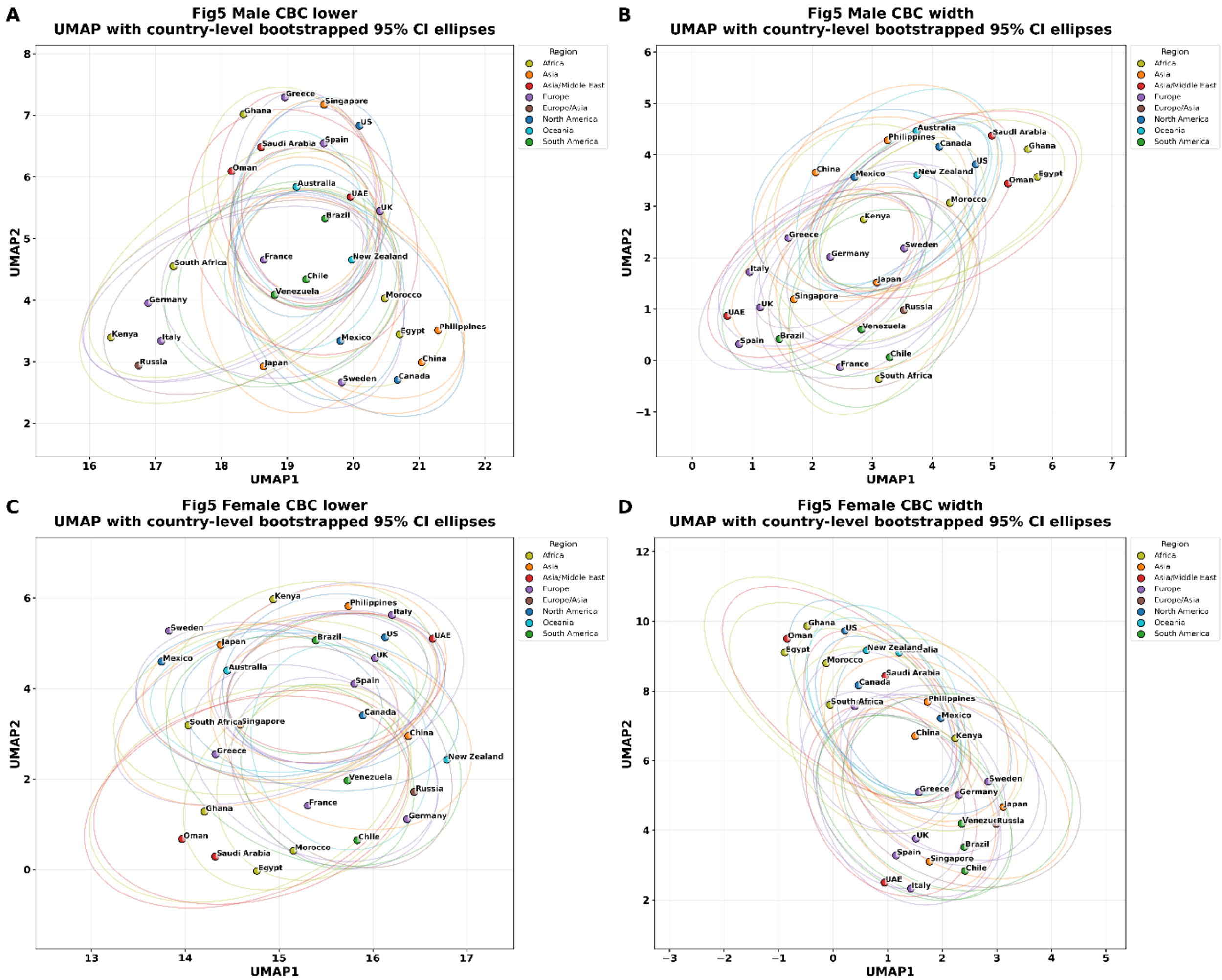

**Figure 5. Robustness of cross-national CBC structure across reference-interval representations and sex with regional 95% confidence ellipses.** UMAP embeddings of countries based on all CBC analytes using alternative representations of published reference intervals. The broad overlapping confidence ellipses indicate no reproducible continent-level separation across 500 bootstraps**.**

(A) Male reference intervals using lower (bottom) limits only.

(B) Male reference-interval widths (upper − lower).

(C) Female reference intervals using lower (bottom) limits only.

(D) Female reference-interval widths (upper − lower).

Countries are coloured by geographic region. Across all representations and both sexes, no stable continent-level clustering pattern is observed. This indicates that the absence of geographic structure is not an artifact of midpoint-based analysis but persists when interval bounds and widths are explicitly considered. UMAP parameters were held constant across panels (see Methods). These analyses rule out the possibility that continent-level structure is masked by summary statistics, boundary selection, or nonlinear projection artifacts. Full analyte-level correlation and permutation importance scores across all panels and reference-interval representations are provided in Supplementary Information.

**Robinson–Foulds tree-distance analysis**

To quantify topological divergence among the six linkage–distance methods, pairwise Robinson–Foulds (RF) distances were computed across 1000 permutations (Figure 6). RF proportions ranged from 0.32 to 1.00 (mean 0.77), indicating high and widespread topological instability across methods (Figure 6). The greatest concordance was observed between magnitude-based metrics: Ward + Euclidean versus Average + Euclidean (mean RF = 0.44) and Ward + Euclidean versus Average + Manhattan (mean RF = 0.64), reflecting their shared sensitivity to absolute CBC magnitudes. By contrast, trees constructed with Average + Mutual Information were topologically orthogonal to all five distance-based trees (mean RF = 0.94-0.95), indicating that the MI-derived topology captures a fundamentally distinct dependence structure from any linear or geometric metric. Across representations and sexes, no two methods consistently produced the same dendrogram topology, confirming that apparent clustering patterns in CBC reference intervals are highly dependent on the choice of distance metric rather than reflecting stable biological signal. This topological instability is consistent with, and independently corroborates, the uniformly high cohesion scores reported in the previous section. When each CBC tree was compared against a geography reference tree (constructed from continent labels), all six methods yielded RF proportions of 0.96–1.00 with no significant departure from the permutation null (all $p > 0.05$), confirming that none of the clustering methods recovered geography-linked topology in CBC reference intervals.

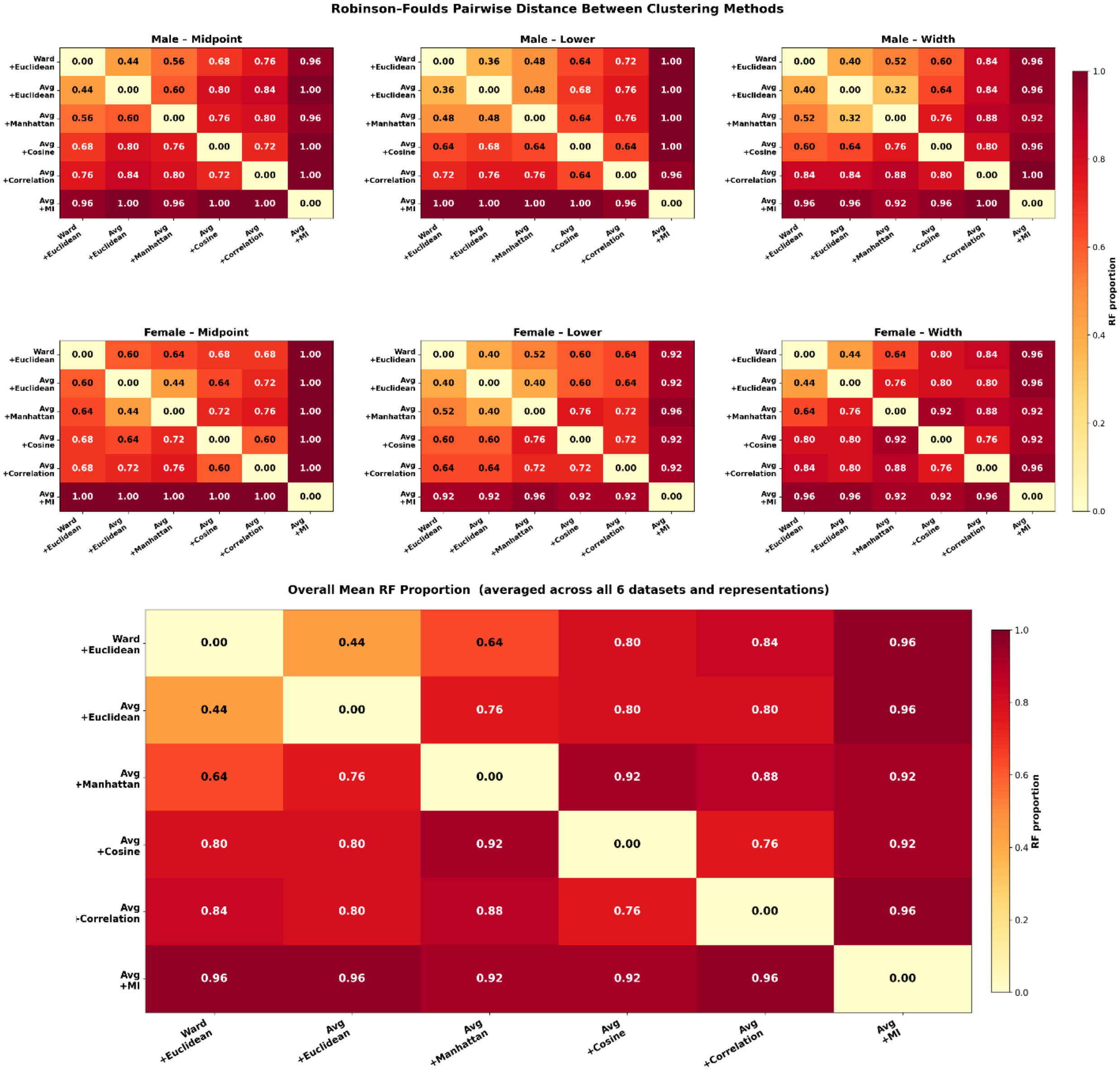

**Figure 6. Robinson–Foulds pairwise distance heatmaps between clustering methods. Each** cell shows the RF proportion (0 = identical topology, 1 = fully divergent) between two linkage–distance combinations. Upper panel: per-dataset results stratified by sex and reference-interval representation (midpoint, lower bound, width). Lower panel: global mean RF proportion averaged across all six datasets and representations. Ward + Euclidean and Average + Euclidean are the most topologically concordant pair (mean RF = 0.44), whereas Average + MI is orthogonal to all distance-based methods, consistent with its capture of non-Euclidean dependence structure. High RF values across all pairs confirm that CBC reference-interval topology is highly metric-dependent and does not reflect a stable biological signal. RF analysis independently confirms the topological instability of clustering across all methods and converges with all other evidence that published reference-interval data, as currently available, do not encode reproducible geography-linked organisation.

## Discussion

### Key Findings and Interpretations

Across all analytic layers—including RI dispersion (Figures 1 and 2), hierarchical clustering (Figure 3), UMAP embeddings (Figure 4), and quantitative benchmarking using the Two-Level Cohesion Score (Tables S2–S5)—CBC reference intervals showed **no** reproducible geography-driven structure. Mutual Information–based clustering occasionally produced weak, localized structure in red-cell indices, particularly MCV and sometimes HGB, but these effects were non-replicating across sexes, feature sets, or distance functions**,** indicating instability rather than a stable population signal. By contrast, BMI produced consistent continent-level organization under the same pipeline (Figure 3C–D; Tables S4–S5), validating that the negative CBC result is an absence of structure and not a modeling limitation.

These findings indicate that CBC reference intervals do not encode a coherent geographical pattern at the continental level. This result should be interpreted as a property of reference-interval standards. While prior work has shown that individual hematologic setpoints can be stable and clinically informative, the present study does not directly assess individual-level physiology. Instead, it demonstrates that aggregating reference intervals across countries fails to yield stable geography-linked structure. Independent haematological studies have reported that intra-individual variability is substantially lower than inter-individual variability, supporting the concept of personalised baselines as clinically relevant anchors. For example, Foy et al. [34] showed that personal hematologic setpoints persist for decades and outperform population-based reference intervals in detecting deviation states. This contrast supports the interpretation that the negative CBC result is not due to a general failure of the network physiology pipeline but rather reflects the absence of reproducible geography-linked structure in the currently available RI datasets. Alternative hypotheses or explanations, including smaller biological effects, individual-level setpoint variation, causal generative mechanisms, epigenetic and psychosocial determinants, nutritional status, altitude, inflammation, environmental exposures, and disease signatures such as anaemia prevalence, to name a few, may also influence CBC distributions but cannot be resolved with reference-interval–level data [35–37].

The present study extends this reasoning to the global scale: if CBC setpoints are intrinsically individual, then aggregating individuals into continents or ethnic blocks will not yield physiologically meaningful reference partitions, precisely what we observed. The SDR visualization (Figure 1) first indicated low cross-national dispersion; clustering instability (Figure 3A–B) confirmed the absence of group structure; the cohesion benchmarking (Tables S2–S5) suggested that even the most sensitive distance metric (MI) could not stabilize geography-based groupings for CBC; and the manifold geometry in UMAP embeddings (Figure 4A–B) showed that no latent spatial separation exists even in a nonlinear representation.

Critically, the success of BMI as a positive case (Figure 3C–D) shows that when geography *does*

drive biological structure, this pipeline detects it. The contrast therefore functions as an internal sensitivity control**,** supporting the conclusion that CBC lacks a continent-level phenotype.

Beyond the empirical findings, the modelling pipeline yields several methodological insights that advance RI research beyond descriptive comparison. First, weak apparent structure occasionally emerged under mutual-information clustering, but these effects consistently disappeared once the feature set or discretization scheme was re-specified, indicating that the signal was method-induced rather than physiologically anchored (Tables S1–S4 in Supplementary Information; Figure 3). Ward+Euclidean served as the most stable baseline configuration (Figure 3A–B), reinforcing that CBC deviations are not variance-cohesive in the way a geography-structured phenotype would be. By contrast, mutual-information distance was the only metric sensitive enough to detect the BMI geography signal (Figure 3C–D; Table S4), confirming that MI operates as a legitimate detector of population structure rather than merely amplifying noise. The UMAP embeddings (Figure 4) function as a nonlinear stress-test: if any underlying manifold separation existed in CBC, it would surface here; its absence therefore validates the negative dendrogram and cohesion results. Together with large effect sizes (Cohen's d = 1.24) and a clinically interpretable enrichment (OR = 6.15; pairs with short distances were approximately 6× likelier to be highly correlated), the strong nonlinear concordance (dCor = 0.74) and low-dimensional geometry (≥87% variance by 4D) indicate that our matrix topology is internally consistent rather than a metric artifact, further supporting the absence of continent-level CBC structure in published RI data rather than modeling limitations.

The Robinson–Foulds (RF) analysis (Figure 6) provides the strongest independent confirmation of the null result in this study, since it is methodologically orthogonal to every prior analytical layer. While cohesion scoring and UMAP embeddings operate on distance-matrix geometry and coordinate spaces, the RF metric topologically assesses whether any two clustering methods converge on which countries belong together? Across all six linkage–distance combinations and all reference-interval representations, mean RF scores were 0.77, confirming that each method produces a largely unique dendrogram topology rather than converging on a shared biological grouping. This finding along with our independent methods confirms the absence of geography-linked structure across clustering coherence, nonlinear embedding geometry, and tree topology. The topological orthogonality of Average + Mutual Information trees to all five distance-based methods further resolves the interpretive challenge posed by MI-based clustering, demonstrating that MI trees capture a fundamentally distinct dependence geometry rather than a noisier version of the same structure. This might partially explain why MI-based clustering occasionally produced weak local groupings that consistently disappeared upon metric re-specification. Comparison of all six CBC trees against a geography reference tree yielded RF proportions of 0.96–1.00 (permutation $p > 0.05$ throughout), confirming that no clustering method regardless of its sensitivity profile recovers geography/continent-level topology in CBC reference intervals.

White blood cell count (WBC) was selected as a representative analyte for detailed visualization because it is among the most universally reported CBC parameters, is commonly sex-agnostic in clinical reference standards, and plays a central role in diagnostic decision-making across health systems. Importantly, multivariate analyses incorporating all CBC analytes similarly failed to reveal consistent geographic clustering, regardless of whether reference intervals were represented by midpoints, bounds, or widths (Figures 2 and 5). This indicates that the observed absence of continent-level structure is not specific to WBC but reflects a broader property of published CBC reference intervals.

Collectively, the lack of separation across all four analytical layers—RI dispersion (Figure 1), hierarchical topology (Figure 3), cohesion benchmarking (Tables S2–S5), and nonlinear manifold structure (Figure 4), constitutes convergent evidence that CBC reference intervals do not encode continent-level structure. In short, the algorithmic result mirrors the physiological one: CBC is not globally structured, whereas BMI is.

**Limitations and Future Directions**

The study's principal limitation is that CBC data are available only as reference intervals, not raw individual-level distributions. This prevents aggregation-level stability from being decomposed into within-population vs cross-population signals**.** Further, a second limitation is heterogeneity of national source authority, although this was explicitly annotated (Figure 1) and its effects were modeled structurally in downstream robustness checks. Source heterogeneity is a limitation of the current global RI landscape but also reflects the real-world complexity and fragmentation that this study aims to characterise. A reduction to fewer countries would substantially limit geographic coverage and preclude robust continent-level inference. Future studies using personalised CBC datasets with analyser metadata, multiomics profiling, environmental covariates, and longitudinal sampling will be required to distinguish true biological homogeneity from structure masked by measurement heterogeneity or RI summarisation. Mutual-information clustering illustrates a third limitation: high sensitivity to discretization choices**,** reinforcing the need for reproducibility safeguards when RI variance is low. Importantly, binning-sensitivity analyses (Supplementary Figures S1–S2) and seeded UMAP embeddings jointly suggest that the absence of geography-linked CBC structure is robust to discretization choices and nonlinear stochastic projection Additional limitations include the absence of age-stratified CBC reference intervals for many countries, which precludes assessment of life-course or developmental effects on reference standards. Moreover, we did not control for environmental or population-level modifiers such as altitude, nutritional status, or anaemia prevalence, as such metadata are not reported alongside available RIs,

An important consideration is the substantial heterogeneity in how CBC reference intervals are established, documented, and updated across countries. Rather than treating this heterogeneity as a confound to be adjusted away, our analysis explicitly captures it as a defining feature of the current global reference-interval landscape. The absence of standardized metadata on analysers, reagents, sampling strategies, or derivation protocols reflects real-world clinical practice, where such information is rarely reported alongside reference intervals. Consequently, our findings do not imply biological uniformity but instead indicate that the available reference-interval data and system themselves lack coherent geographic or population-level structure. This structural heterogeneity rather than hidden biological variation appears to dominate cross-national differences in CBC reference intervals. Notably, the high prevalence of sex-agnostic reference intervals for several CBC analytes (Table S5) reflects current clinical practice and further underscores that the absence of geographic clustering is not an artifact of sex stratification choices.

Future work should address these gaps by using large-scale individual-level CBC datasets with a diverse and multi-institutional cohort, as well as incorporating longitudinal measurements to reconstruct personal hematologic trajectories. In addition, prospective analyses can integrate genomic, environmental, and demographic variables to model the "individual baseline" as complex systems and make mechanistic inferences. We propose that formalizing RI evolution toward adaptive or personalised reference systems, rather than geographically universal ones could benefit translatability to predictive medicine frameworks.

A concrete next step would involve multi-country, prospective recalibration of CBC reference intervals using harmonized sampling protocols across accredited laboratories, paired with standardized metadata on social determinants of health and demographic factors including age, sex, analytical platform, and calibration procedures. Clinically, this could be implemented through multi-step integration of longitudinal CBC monitoring within laboratory information systems, enabling personal baseline tracking in primary care, including rural and underserved settings. Such designs would allow validation of individualized RIs against clinical outcomes and support translation into routine diagnostic workflows. Importantly, such translatable frameworks could be deployed using low-cost, routinely collected CBC data, making equitable personalised diagnostics accessible beyond tertiary centres and supporting marginalized community-oriented public health.

Monte Carlo power simulations with permutation-based significance testing ($\alpha = 0.05$) showed limited sensitivity to continent-level shifts when using midpoint-based reference intervals. Across analytes, power remained near nominal levels for shifts $\Delta \leq 1$ SD and increased only modestly for large shifts ($\Delta \approx 2$ SD), particularly in the female subset where continent group sizes were small. Accordingly, effects smaller than 1.5–2 SD would be difficult to detect under the current design (Figure S3 in Supplementary Information). This in silico power analysis indicates that the absence of a detectable geography-linked pattern is unlikely to reflect insufficient statistical power but rather reflects either small effects or lack of structure at the resolution afforded by available RI systems.

Finally, although random seeds were fixed and all parameters documented, methods such as UMAP and permutation importance remain inherently stochastic. Future work may therefore benefit from deterministic manifold-learning approaches or fully Bayesian embedding frameworks to further improve run-to-run stability, without altering the core qualitative conclusions reported here.

Our findings carry important implications for accessible, affordable, and precision diagnostics including underserved and low resource community care settings. This could include harmonized international reporting standards for RIs, mandatory disclosure of de-identified (anonymized) metadata on RI derivation (e.g., population source, analytical platform, and update year), and phased integration of longitudinal individual baselines within accredited laboratory information systems. Highlighting that CBC reference intervals lack stable geographic structure reframes population reference intervals from region-specific correction factors toward individualized longitudinal baselines, emphasizing that true diagnostic precision lies in tracking intra-individual trends over time. These position CBC-derived circulatory biomarkers as translatable, low-cost (affordable) substrates for longitudinal monitoring and equitable access to personalised care.

## Conclusions

Our findings suggest that CBC reference intervals do not exhibit reproducible continent-level or geography-linked structure in the compiled global datasets. Rather, they are consistent with a paradigm of individualized hematologic setpoints, in which personalised longitudinal stability, rather than population averages, anchors clinical interpretability. In contrast, BMI retains robust geography-linked structure, validating the analytical framework and indicating that the absence of CBC clustering reflects properties of published reference-interval systems rather than an analytic artefact.

Importantly, the external validity of these conclusions is constrained by heterogeneity in data sources, reference-interval derivation practices, and the absence of individual-level measurements. Accordingly, these findings support, within the limits of aggregated reference-interval data, a transition from population-based to individual-based reference systems, with CBC values interpreted against personal baselines rather than presumed universal norms, as a potential pathway toward precision diagnostics and diagnostic equity. Personalised reference systems should be interpreted here as a translational implication together with previous evidence [34, 38, 39], rather than as a direct conclusion. Future development of dynamic, data-adaptive RI frameworks will require longitudinal and multimodal integration, eventually recalibrating precision hematology around the “human”, the individual, in personalised medicine.

**Competing Interests**: One author holds editorial board positions; however, all editorial roles were fully recused from the handling, review, and decision process for this manuscript, which was managed independently in accordance with journal policy. No other conflicts of interest exist.

**Funding Statement**: Not applicable.

**Data and Code Availability:** All CBC reference-interval sources, country-level datasets, and analysis scripts used in this study are documented in Appendix A and Supplementary Materials. Computational analyses were performed using Python, and all parameter settings required to reproduce figures and results are explicitly reported in the Methods. Due to the use of published reference intervals rather than individual-level data, no raw patient data are included. The original codes are provided in the following repository: https://github.com/kunlin889797/Proximity-and-Similarity-Analysis-of-Complete-Blood-Count-

**Author contributions:** KW: Conceptualisation, data curation, formal analysis, investigation, methodology, resources, software, validation, visualisation, writing, writing review and editing. AU: Formal analysis, investigation, methodology, software, validation, writing, writing review and editing. HZ: Conceptualisation, project administration, resources, supervision, writing review and editing.